\documentclass[10pt,a4]{amsart}

\usepackage{amssymb}
\usepackage{amsmath}
\usepackage{verbatim}
\usepackage{enumerate}
\usepackage{hyperref}
\usepackage{bm}   
\usepackage{extarrows}
\usepackage{longtable}
\usepackage{amsthm}
\usepackage{amscd}

\newcommand{\RR}{\mathbb{R}}
\newcommand{\CC}{\mathbb{C}}
\newcommand{\ZZ}{\mathbb{Z}}
\newcommand{\sa}{\mathsf{a}}
\newcommand{\sabar}{\bar{\mathsf{a}}}
\newcommand{\A}{a}
\newcommand{\Abar}{\bar{a}}
\newcommand{\dd}{\partial}
\newcommand{\del}{\bm\partial}
\newcommand{\delb}{\bm{\bar\partial}}
\newcommand{\ddel}{\partial}
\newcommand{\ddelb}{\bar\partial}
\newcommand{\tot}{\mathrm{tot}}
\newcommand{\simeom}{\underset{\mathrm{e.o.m.}}{\sim}}
\newcommand{\lan}{\left\langle}
\newcommand{\ran}{\right\rangle}
\newcommand{\mc}{\mathcal}
\newcommand{\mr}{\mathrm}
\newcommand{\bb}{\mathbb}
\newcommand{\s}{\mathsf}
\newcommand{\ra}{\rightarrow}
\newcommand{\hra}{\hookrightarrow}
\newcommand{\til}{\widetilde}
\newcommand{\OO}{\mathcal{O}}
\newcommand{\cF}{\mathcal{F}}
\newcommand{\reg}{\mathrm{reg.}}

\newcommand{\mb}{\mathbf}
\newcommand{\Fbun}{\underline{\mathbb{F}}}
\newcommand{\F}{\mathsf{\underline{F}}}

\newcommand{\wh}{\widehat}

\theoremstyle{definition}
\newtheorem{remark}{Remark}[section]
\newtheorem{example}[remark]{Example}
\allowdisplaybreaks

\begin{document}

\title[Abelian $BF$ theory as a conformal field theory]{Two-dimensional abelian $BF$ theory in Lorenz gauge as a twisted $\mathcal{N}=(2,2)$ superconformal field theory}

\begin{abstract}
We study  the two-dimensional topological 
abelian $BF$ theory in the Lorenz gauge and, surprisingly, we find 
that the gauged-fixed theory is a free type B twisted $\mc{N}=(2,2)$ superconformal theory with
odd linear target space, with the ghost field $c$ being the pullback
of the linear holomorphic coordinate on the target. The 
$Q_\mr{BRST}$ 
of the gauge-fixed theory equals
the total $Q$ of type B twisted theory. This unexpected identification 
of two different theories opens a way for nontrivial deformations of both of these theories.

\end{abstract}

\author{Andrey S. Losev}
\address{Federal Science Centre ``Science Research Institute of System Analysis at Russian Science Academy''
(GNU FNC NIISI RAN), Moscow, Russia}
\address{National Research University Higher School of Economics (HSE), Moscow, Russia}
\address{Moscow Institute of Physics and Technology (MIPT), Dolgoprudnyi, Russia}
\address{Wu Key Lab, USTC}
\email{aslosev2 @gmail.com}

\author{Pavel Mnev}
\address{
University of Notre Dame 
}
\address{
St. Petersburg Department of V. A. Steklov Institute of Mathematics of the Russian Academy of Sciences
}
\email{pmnev @nd.edu}

\author{Donald R. Youmans}
\address{
Universit\'e de Gen\`eve
}
\email{Donald.Youmans @unige.ch}

\thanks{The work of A. L. was accomplished in GNU
FNC NIISI RAN program No.6, theme 36.20; the work of A. L. was supported by the RFBR grant 15-01-09242a.
 P. M. acknowledges partial support of RFBR Grant No. 17-01-00283a. Research of D. Y. was supported by the Grant 165666 and the NCCR SwissMAP of the Swiss National Science Foundation.}


\maketitle

\setcounter{tocdepth}{3} 
\tableofcontents

\section*{
Foreword
}
The note is designed as a self-contained exposition, 
with relevant 
background on (super)conformal field theory included in the text for the reader's convenience. We refer to the sources \cite{Polchinski, Witten_mirror, Hori} for details.

For the readers who are well acquainted with superconformal field theory, we suggest to look at the formula (\ref{S via complex fields}) and then read the subsections \ref{sec: anticipation} and \ref{sec: dictionary} for  
the quick 
gist of the story.


\subsection*{Acknowledgements} We would like to thank Anton Alekseev, Alberto S. Cattaneo and Stephan Stolz for stimulating discussions.

\section{Classical theory}

We consider the abelian $BF$ theory on a two-dimensional oriented surface $\Sigma$ 
defined classically by the action functional 
\begin{equation}\label{S_0}
S_0=\int_\Sigma B\, dA
\end{equation}
where the fields are a $1$-form $A$ and a $0$-form $B$ on $\Sigma$; $d$ is the de Rham operator. The equations of motion read $dA=0$, $dB=0$ and the theory has gauge symmetry ${A\mapsto A+d\alpha}, B\mapsto B$ with the 
$0$-form $\alpha$ being 
the generator of the gauge transformation.

\subsection{Gauge-fixed model in BRST formalism}
We impose the Lorenz gauge condition $d^*A=0$ where $d^*=- *d*$ 
is the Hodge dual of the de Rham operator associated to a choice of a Riemannian metric $g$ on $\Sigma$, 
with $*$ the Hodge star. The corresponding gauge-fixed action (the Faddeev-Popov action) is:
\begin{equation}\label{S}
S=\int_\Sigma B\, dA+ \lambda\, d*A+b\,d*dc 
\end{equation}
with scalar field $\lambda$ the Lagrange multiplier imposing the Lorenz gauge condition and $b,c$ the Faddev-Popov ghosts -- the odd scalar fields. Thus, the space of BRST fields of the model is:
\begin{equation}\label{BRST fields}
\mathcal{F}=\underbrace{\Omega^1}_{A}\oplus \underbrace{\Omega^0}_{B}\oplus \underbrace{\Omega^0}_{\lambda}\oplus \underbrace{\Pi\Omega^0}_{b}\oplus \underbrace{\Pi \Omega^0}_{c}
\end{equation}
with $\Pi$ the parity reversal symbol. Equivalently, it is the space of sections $\mathcal{F}=\Gamma(\Sigma,\F)$ of the super vector bundle 
\begin{equation}\label{BRST bundle}
\F=\underbrace{T^*\Sigma}_{A}\oplus \underbrace{\underline{\RR^{2}}}_{B,\lambda}\oplus \underbrace{\underline{\Pi \RR^2}}_{b,c}
\end{equation}
over $\Sigma$, with 
last two terms the trivial even and odd rank $2$ bundles.\footnote{ 
One can enhance the $\ZZ_2$-grading 
on fields to a $\ZZ$-grading by the ``ghost number'', where $b$, $c$ are assigned degrees $-1$ and $+1$, respectively, and $A,B,\lambda$ are assigned degree $0$.
}
The BRST operator 
acts as 
\begin{equation*}
Q:\qquad  A\mapsto dc,\quad b\mapsto\lambda, \quad B,c,\lambda\mapsto 0
\end{equation*}
One clearly has $Q^2=0$ and $Q(S)=0$ -- the gauge-invariance of the action. Also, the gauge-fixed action differs from $S_0$ by a $Q$-exact term:
\begin{equation}\label{S=S_0+Q Psi}
S=S_0+Q(\Psi)
\end{equation}
with 
\begin{equation}\label{Psi}
\Psi=\int_\Sigma b\, d * A
\end{equation}
the gauge-fixing fermion.

Equations of motion for the gauge-fixed action (\ref{S}) read
\begin{equation*}
dA=0,\quad d*A=0,\quad dB-*d\lambda=0,\quad \Delta b=0,\quad \Delta c=0
\end{equation*}
with $\Delta=*d*d$ the Laplacian acting on functions on $\Sigma$.

\subsection{Rewriting the action in terms of complex fields, conformal invariance}
It is convenient to split the $1$-form field $A$ into its $(1,0)$ and $(0,1)$-components $A=\sa+\sabar$, 
where the splitting $\Omega^1=\Omega^{1,0}\oplus \Omega^{0,1}$ is inferred from the complex structure on $\Sigma$ compatible with the chosen metric (in particular, $\Omega^{1,0}$ and $\Omega^{0,1}$ are the $-i$ and $i$-eigenspaces for the Hodge star). We also combine the real scalar fields $B,\lambda$  into a complex scalar field $\gamma:=\frac12 (\lambda+iB)$ with conjugate $\bar\gamma=\frac12(\lambda-iB)$. 
Action (\ref{S}) can then be written as
\begin{equation}\label{S via complex fields}
S=2i\int_\Sigma -\gamma\, \delb \sa + \bar\gamma\, \del \sabar + b\, \del \delb\, c
\end{equation}
with $\del,\delb$  the holomorphic and anti-holomorphic Dolbeault operators (written as $\del=dz\frac{\dd}{\dd z},\delb=d\bar{z}\frac{\dd}{\dd \bar{z}}$ in local complex coordinates $z,\bar{z}$; we reserve the non-boldface symbols $\ddel=\frac{\dd}{\dd z},\ddelb=\frac{\dd}{\dd \bar{z}}$ for the partial derivatives themselves). Written in this form, the action is manifestly
dependent only on the complex structure induced by the metric $g$, i.e. only on the conformal class of the metric $g$ modulo Weyl transformations $g\sim \Omega\cdot g$ with $\Omega$ any positive function on $\Sigma$. Thus the gauge-fixed abelian $BF$ theory is conformal. The BRST operator $Q$ written in terms of the new fields reads:
\begin{equation}\label{Q in complex fields}
Q:\qquad \sa\mapsto \del c,\quad \sabar\mapsto \delb c,\quad b\mapsto \gamma+\bar\gamma,\quad\quad \gamma,\bar\gamma,c\mapsto 0
\end{equation}
and the equations of motion are:
\begin{equation}\label{equations of motion}
\delb \sa=0, \quad \delb\gamma=0,\quad \del \sabar=0,\quad \del \bar\gamma=0,\quad \del\delb\, b=0,\quad \del\delb\, c=0
\end{equation}

Note that fields $\gamma,\bar\gamma$ are more adapted to the action and the equations of motion, whereas fields $B,\lambda$ are more adapted to the BRST operator.

\begin{remark}
Using local coordinates $x^1,x^2$ on $\Sigma$ (such that $g$ is in the conformal class of $(dx^1)^2+(dx^2)^2$) and the corresponding complex coordinates $z=x^1+ix^2$, $\bar{z}=x^1-ix^2$, we can write $\sa=dz\, \A$, $\sabar=d\bar{z}\, \Abar$ with $\A,\Abar$ scalars. Then the action (\ref{S via complex fields}) reads
\begin{equation}\label{S coord}
S=4\int d^2x\;(\gamma \ddelb \A +\bar\gamma \ddel \Abar + b\,\ddel\ddelb c)
\end{equation}
with $d^2 x=dx^1 dx^2=\frac{i}{2}dz\, d\bar{z}$ the coordinate area form. In our conventions, for the fields $\A,\Abar$ one gets a sign in BRST transformations, $Q:\; \A\mapsto -\ddel c, \;\Abar\mapsto -\ddelb c$. 

\end{remark}

The BRST symmetry  
defines, via the Noether theorem, a current 
\begin{equation}\label{J BRST}
J_\tot=
2i (\gamma \,\del c - \bar\gamma\,\delb c) = 2i\bigg(dz\,\underbrace{\gamma\,\ddel c}_{=:J}-d\bar{z}\,\underbrace{\bar\gamma\,\ddelb c}_{=:\bar{J}}\bigg)
\end{equation}
It is conserved modulo equations of motion:  $dJ_\tot\underset{\mathrm{e.o.m.}}{\sim}0$. 
In fact, one has a stronger statement that both chiral parts of the current are conserved independently:  $\ddelb J\simeom 0$ and $\ddel \bar{J}\simeom 0$.

\subsubsection{Abelian $BF$ theory as a twisted superconformal field theory: 
an anticipation
} \label{sec: anticipation}
It is remarkable -- see section \ref{sec: U(1) current and untwist} for details -- that action (\ref{S via complex fields}) is a free 
type B twisted $\mc{N}=(2,2)$ superconformal theory where the
parity of the fields is changed (so scalars are fermions
 while the first order systems are constructed with bosons).
The ``holomorphic field'' is just the Faddev-Popov ghost.
As we will show later the $Q_\mr{BRST}$ becomes a sum of
two scalar charges, as
usual in the twisted theory; their currents have changed the dimension
from $3/2$ to $1$ in both holomorphic and antiholomorphic sectors.
This unexpected property allows to pose questions in $BF$ theory
 that where prohibited by the naive understanding of allowed correlators --
in particular, 
it is possible to study correlators of some
gauge non-invariant observables, like the superpartner (BRST-primitive) of the
energy-momentum tensor (see below). On the other hand non-abelian
$BF$ theory in Lorenz gauge may serve as an example of a new conformal
field theory (this question is clear classically and we will return
to this important question on the quantum level in subsequent work).

\subsection{Stress-energy tensor and its BRST-primitive}
The stress-energy tensor is defined via the variational derivative of the action (\ref{S}) with respect to metric. Explicitly, in a coordinate chart on $\Sigma$, one defines $T_{\mu\nu}$ via 
\begin{equation}\label{T as dS/dg}
\delta_g S=-\int_\Sigma \sqrt{\det g}\, d^2 x\, T_{\mu\nu} \delta g^{\mu\nu}
\end{equation}
where the left hand side is the variation w.r.t. the metric $g$; indices $\mu,\nu$ take values in $\{1,2\}$ or $\{z,\bar{z}\}$. The total stress-energy tensor $T_\tot=T_{\mu\nu}\,dx^\mu\cdot dx^\nu$ is 
a section of the symmetric square of the cotangent bundle of $\Sigma$; the dot stands for the symmetric tensor product in $\mathrm{Sym}^\bullet T^*\Sigma$. Note that $S$ depends on the metric only via the dependence of the gauge-fixing fermion $\Psi$ on the metric, entering via the Hodge star. Thus, from (\ref{S=S_0+Q Psi}) we have that the components of the stress-energy tensor are exact w.r.t. the BRST operator, 
\begin{equation*}
T_{\mu\nu}=QG_{\mu\nu}
\end{equation*}
where $G_{\mu\nu}$ is defined, similarly to (\ref{T as dS/dg}), via 
\begin{equation*}
\delta_g \Psi=-\int_\Sigma \sqrt{\det g}\, d^2 x\, G_{\mu\nu} \delta g^{\mu\nu}
\end{equation*}

Explicitly, in holomorphic coordinates $z,\bar{z}$, one obtains\footnote{
Here is the computation in local coordinates: Hodge star acts on $1$-forms via $*dx^\mu=\sqrt{g}\,g^{\mu\nu}\epsilon_{\nu\rho}dx^\rho$ with $\epsilon_{\nu\rho}$ the Levi-Civita symbol and $\sqrt{g}$ the shorthand notation for $\sqrt{\det g}$. Variation w.r.t. the metric is thus $\delta_g *dx^\mu=\sqrt{g}(-\frac12 g_{\alpha\beta}\delta g^{\alpha\beta}\,g^{\mu\nu}+\delta g^{\mu\nu})\epsilon_{\nu\rho}dx^\rho$. Next, we use this to compute the variation of 
the gauge-fixing fermion:
$\delta_g \Psi=\int db\wedge \delta_g *dx^\mu\,A_\mu=\int \sqrt{g}\, db\wedge (-\frac12 g_{\alpha\beta}g^{\mu\nu}+\delta^\mu_\alpha\delta^\nu_\beta)\epsilon_{\nu\rho}dx^\rho A_\mu \delta g^{\alpha\beta}$. Note that the coefficient of $\sqrt{g}\,\delta g^{\alpha\beta}$ in the integrand is manifestly invariant under Weyl transformations $g\mapsto \Omega\cdot g$. In particular, we can compute this coefficient in holomorphic coordinates $z,\bar{z}$, using the standard metric $g=dz\cdot d\bar{z}$. One obtains
$\delta_g \Psi=-\int\frac{i}{2}dz\wedge d\bar{z}\,(\ddel b\, \A\, \delta g^{zz}+\ddelb b\, \Abar\,\delta g^{\bar{z}\bar{z}})$; reading off the coefficients of the variation of the metric, we obtain (\ref{G}).
} 
\begin{equation}\label{G}
G_\tot=(dz)^2 \underbrace{\A\,\ddel b}_{G_{zz}=:G}+ (d\bar{z})^2 \underbrace{\Abar\,\ddelb b}_{G_{\bar{z}\bar{z}}=:\bar{G}}=
\sa\cdot \del b + \sabar\cdot \delb b
\end{equation}
and 
\begin{multline*}
T_\tot=Q G_\tot = 
(dz)^2\underbrace{(-\ddel c\;\ddel b+\A\,\ddel \lambda)}_{T_{zz}=:T}+
(d\bar{z})^2\underbrace{(-\ddelb c\;\ddelb b+\Abar\,\ddelb \lambda)}_{T_{\bar{z}\bar{z}}=:\bar{T}}
\\
=(\del c\cdot \del b+\sa\cdot \del \lambda) + (\delb c\cdot \delb b+\sabar\cdot \delb \lambda)
\end{multline*}
The components $G_{z\bar{z}}$, $T_{z\bar{z}}$ vanish (equivalently, the traces $G^\mu_{\;\;\mu}$, $T^\mu_{\;\;\mu}$ vanish), which is a manifestation of the conformal invariance of the theory. The stress-energy tensor and its primitive are conserved modulo equations of motion:
\begin{equation*}
\ddelb G\simeom 0, \quad \ddel \bar{G}\simeom 0;\qquad \ddelb T\simeom 0,\quad \ddel\bar{T}\simeom 0
\end{equation*}

\section{Quantum abelian $BF$ theory as a conformal field theory}   

From now on we specialize to the case of the surface $\Sigma$ being the plane $\RR^2=\CC$ with coordinates $x^1,x^2$ (or the complex coordinate $z=x^1+ix^2$ and its conjugate $\bar{z}=x^1-ix^2$), endowed with the standard Euclidean metric $g=(dx^1)^2+(dx^2)^2=dz\cdot d\bar{z}$.

\subsection{Correlation functions}
We are interested in studying the normalized correlation functions 
\begin{equation}\label{correlator}
\lan \Phi_1(z_1)\cdots \Phi_n(z_n) \ran := \frac{1}{Z} \int_\cF e^{-\frac{1}{4\pi}S}\; \Phi_1(z_1)\cdots \Phi_n(z_n)
\end{equation}
Here:
\begin{itemize}
\item $\Phi_1(z_1),\cdots ,\Phi_n(z_n)$ are composite fields\footnote{
We are using this terminology to emphasize the distinction between the ``basic'' BRST fields (\ref{BRST fields}) and the objects that can be used as decorations of punctures on $\Sigma$ when calculating correlation functions. We call the latter the composite fields. Another possible term is ``observables'', or ``$0$-observables'' (though often one reserves the word ``observable'' only for the $Q$-closed expressions in fields).
} 
-- polynomials in the fields $a, \bar{a}, B, \lambda, b, c$ and their derivatives of arbitrary order, evaluated at pairwise distinct points $z_1,\ldots,z_n\in\CC$.
\item $S$ is the action 
(\ref{S coord}).
\item The normalization factor $Z:=\int_\cF e^{-\frac{1}{4\pi}S}$ is the partition function. The r.h.s. of (\ref{correlator}) is a ratio of path integrals over $\cF$ which individually need a regularization (both ultraviolet and infrared) to be defined. However the ratio is independent of the regularization.
\item The normalization factor $\frac{1}{4\pi}$ in the exponential in the r.h.s. of (\ref{correlator}) is introduced to have a convenient normalization of propagators.
\end{itemize}

Since the action $S$ is quadratic, the theory is free and the correlators are given by Wick's lemma with the following basic propagators\footnote{
One constructs the propagator $\lan c(z)\,b(w) \ran$ from the action (\ref{S coord}) as the Green's function (the integral kernel of the inverse operator) for the operator $\frac{1}{4\pi}\Delta$. Similarly, $\lan a(w)\, \gamma(z) \ran$ is the Green's function for $\frac{1}{\pi}\ddelb$. 
We implicitly fix the zero-mode for the Dolbeault operator by requiring that the fields $a,\gamma,\bar{a},\bar\gamma$ vanish at infinity; in other words we are considering the theory on the compactified plane $\bar\CC=\CC \bb{P}^1$ relative to the point $\{\infty\}$. To fix the zero-mode of the Laplacian we do an infrared regularization by replacing the pair $(\CC \bb{P}^1,\{\infty\})$ with a disk of large radius $R$ relative to the boundary, i.e., we impose the Dirichlet boundary conditions on $b,c$. The constant shift $C$ in the propagator $\lan c(w)\, b(z) \ran $ depends on the infrared cut-off, $C=-2\log R$.
}:
\begin{equation}\label{propagators}
\lan c(w)\, b(z) \ran = 2 \log |w-z|+C,\quad
\lan a(w)\, \gamma(z) \ran = \frac{1}{w-z}, \quad
\lan \bar{a}(w)\, \bar\gamma(z) \ran = \frac{1}{\bar{w}-\bar{z}}
\end{equation}
Propagators for all other pairs of fields from the set $\{a,\bar{a},\gamma,\bar\gamma,b,c\}$ vanish. In terms of fields $A,B,\lambda$ this implies
\begin{equation}\label{propagators AB, Alambda}
\lan A(w)\, B(z)\ran = 2\,d_w \arg(w-z),\quad \lan A(w)\,\lambda(z) \ran = 2\,d_w\log |w-z| 
\end{equation}


When constructing the correlators  (\ref{correlator}) for composite fields by Wick's lemma, we do not allow matchings of two basic fields in the same composite field $\Phi_i$ (which would have led to an ill-defined expression) -- this corresponds to the assumption that the composite fields $\Phi_i$ are normally ordered.

Note also that if one of the fields $\Phi_i$ vanishes modulo equations of motion (\ref{equations of motion}), the correlator (\ref{correlator}) vanishes identically.

\subsection{The space of composite fields}
One can formalize the notion of a composite field by considering the symmetric powers 
of the jet bundle of $\F^*_\CC=\CC\otimes \F^*$ -- the complexified bundle dual to the bundle of
BRST fields, and then taking a quotient by the ideal generated by the equations of motion (\ref{equations of motion}) and their derivatives: 
\begin{equation*}
\Fbun:=\mathrm{Sym}^\bullet \mathrm{Jet}\,\F^*_\CC/\mr{e.o.m.}
\end{equation*} 
It is a graded complex vector bundle over $\Sigma$ with $\bb{Z}$-grading given by the ghost number (by assigning degree $+1$ to $c$ and degree $-1$ to $b$ and zero to all other basic fields). Thus, a composite field $\Phi(z)$, regarded modulo equations of motion, is an element of the fiber $\bb{F}_z$. 
Fibers of $\Fbun$ are differential graded commutative algebras, with the differential given by the BRST operator $Q$. 
The $n$-point correlator (\ref{correlator}) can then be regarded as bundle morphism from $\iota^* (\Fbun\boxtimes \cdots\boxtimes \Fbun) $ to the trivial line bundle over $\mr{Conf}_n(\Sigma)$ -- the (open) configuration space of $n$ pairwise distinct ordered points on $\Sigma$; here $\iota: \mr{Conf}_n(\Sigma)\hra \Sigma^{\times n}$ is the tautological inclusion.

Explicitly, the space of composite fields $\bb{F}_z$ is a free graded commutative algebra generated by the fields
\begin{multline}\label{F_z generators}
b,\; c;\; 
\underbrace{\{\ddel^k b\}_{k\geq 1},\; \{\ddel^k c\}_{k\geq 1},\; \{\ddel^k a\}_{k\geq 0},\; \{\ddel^k \gamma\}_{k\geq 0}}_{\mr{holomorphic\;sector}};\; \\
\underbrace{\{\ddelb^k b\}_{k\geq 1},\; \{\ddelb^k c\}_{k\geq 1},\; \{\ddelb^k \bar{a}\}_{k\geq 0},\; \{\ddelb^k \bar\gamma\}_{k\geq 0}}_{\mr{anti-holomorphic\; sector}}
\end{multline}
The BRST differential $Q$ is a derivation defined on the generators by (\ref{Q in complex fields}) together with the rule $Q(\mc{D}\phi)=
\mc{D}Q(\phi)$ for 
any differential operator $\mc{D}=\dd^k\ddelb^l$
and $\phi$ a basic field, and then extended to the whole $\bb{F}_z$ by Leibniz identity.

The cohomology of $Q$ acting on $\bb{F}_z$ is calculated straightforwardly and yields the subalgebra of $\bb{F}_z$ generated by fields $B=
\frac{\gamma-\bar\gamma}{i}
$ and $c$ (but not their derivatives):\footnote{
Here is the calculation: denote by $Y$ the linear span of the generators (\ref{F_z generators}). Note that, by freeness of the theory, $Q$ acts on $Y$ as a differential and $Q$ on $\bb{F}_z$ is the extension of this action by Leibniz identity. Thus  $H^\bullet_Q(\bb{F}_z)=H^\bullet(\mr{Sym}\,Y)=\mr{Sym}\,H^\bullet (Y)$. To compute $H^\bullet(Y)$, notice that $Q$ maps $\dd^{k-1} a\mapsto -\dd^{k}c$, $\ddelb^{k-1} \bar{a}\mapsto -\ddelb^{k}c$, $\dd^k b\mapsto  \dd^k\gamma $, $\ddelb^k b\mapsto  \ddelb^k\bar\gamma$ for all $k\geq 1$. Thus we can remove the acyclic subcomplex spanned by these generators out of $Y$ and we have $H^\bullet(Y)=H^\bullet\mr{Span}(b,c,\gamma,\bar\gamma)=H^\bullet\mr{Span}(b,c,B,\lambda)$. Finally, since $Q$ maps $b\mapsto \lambda$ and vanishes on $B,c$, we have $H^\bullet(Y)=\mr{Span}(B,c)$ and hence $H^\bullet_Q(\bb{F}_z)=\mr{Sym}\,\mr{Span}(B,c)$.
}
\begin{equation}\label{Q coh}
\bb{O}_z:=H^\bullet_Q (\bb{F}_z)=\CC[B,c]
\end{equation}
We denote the $Q$-cohomology by $\bb{O}_z$ (for ``observables''). Note that it is concentrated in degrees $0$ and $1$ only, since polynomials in $c$ have degree at most $1$. However, one can consider an $N$-component abelian $BF$ theory, i.e., $N$ non-interacting copies of the theory; in other words, one replaces the fiber (\ref{BRST bundle}) of the bundle of BRST fields with  
\begin{equation}\label{N-component BRST fiber}
\F\mapsto \F^{[N]}:=\F\otimes \RR^N
\end{equation} 
(we will use the superscript $[N]$ when we want to emphasize that we work with $N$-component theory). 
In this case, there are $N$ odd generators $c^j$ and the polynomials in them can have degree up to $N$ and thus the cohomology $\bb{O}_z^{[N]}:=H^\bullet_Q(\bb{F}_z^{[N]})=\CC[B_1,\ldots,B_N,c^1,\ldots,c^N]$ is spread across degrees $0,1,\ldots,N$.

One can geometrically interpret the space of observables as the space of polyvector fields 
\begin{equation}\label{observables as polyvectors}
\bb{O}_z^{[N]}=T_\mr{poly}(\Pi V )
\end{equation} 
on the odd space $\Pi V$ where $V=\RR^N$ is the space of coefficents in the $N$-component theory. Here the ghosts $c^j$ are interpreted as coordinates on the base $\Pi V$ and $B_j=\frac{\dd}{\dd c^j}\in \Pi T_c(\Pi V)$ are interpreted as tangent vectors.\footnote{In terms of $\ZZ$-grading: $\bb{O}_z^{[N]}=T_\mr{poly}(V[1])$ -- sections of the symmetric 
powers of the shifted tangent bundle $T[-1](V[1])$ over $V[1]$, or, equivalently, functions on the graded space $T^*[1]V[1]$.}

\subsection{Operator product expansions}

We are interested in analyzing the singular part of the asymptotics  
of the correlator 
\begin{equation}\label{OPE corr}
\lan \Phi_1(w) \Phi_2(z) \phi_1(z_1)\cdots \phi_n(z_n)\ran 
\end{equation}  
as the  point $w$ approaches $z$, with $\phi_1(z_1),\ldots,\phi_n(z_n)$ being the ``test fields''.  This asymptotics is controlled by the operator product expansion (OPE) of the fields $\Phi_1$ and $\Phi_2$ which is an expression of the form
\begin{equation}\label{OPE}
\Phi_1(w)\,\Phi_2(z) \sim \sum_{j=1}^p f_j(w-z) \til\Phi_j(z)+\mr{reg.}
\end{equation}
with $\til\Phi_{j}$ some fields and $f_j$ some singular coefficient functions, typically a product of negative powers of $(w-z)$ and $(\bar{w}-\bar{z})$ and can also contain $\log(w-z)$ and $\log(\bar{w}-\bar{z})$; $\mr{reg.}$ stands for terms which are regular 
as $w\ra z$; the number $p$ of singular terms depends on $\Phi_1,\Phi_2$. Thus, (\ref{OPE}) means that in the correlator (\ref{OPE corr}) one can replace the first two fields with the expression on the r.h.s. of (\ref{OPE}), reducing the number of points by one. 

\begin{example}
For instance, we have 
$$a(w)\,\gamma(z)\sim \frac{1}{w-z}+(a\gamma)_z+(w-z)\cdot (\dd a\,\gamma)_z+\frac12 (w-z)^2\cdot  (\ddel^2 a\,\gamma)_z+\cdots \quad \sim
\frac{1}{w-z}+\mr{reg.}$$
and 
$$ a(w)\,\bar\gamma(z) \sim (a\bar\gamma)_z+(w-z)\cdot (\dd a\,\bar\gamma)_z+\frac12 (w-z)^2\cdot  (\ddel^2 a\,\bar\gamma)_z+\cdots\quad \sim \mr{reg.}$$
For brevity we put the point where the field is evaluated as a subscript (i.e. $\Phi(z)=\Phi_z$). Note that the OPE $a(w)\bar\gamma(z)$ is purely regular: correlators containing this pair of fields have a well-defined limit as $w\ra z$, whereas a correlator containing $a(w)\gamma(z)$ will have a first order pole as $w\ra z$. 
\end{example}

Generally, since we are dealing with a free theory, the OPE $\Phi_1(w)\,\Phi_2(z)$ for two generic composite fields is constructed by Wick's lemma. In particular, for $\Phi_1,\Phi_2$ two monomials in (derivatives of) basic fields, the recipe for OPE is as follows. We consider all partial matchings of basic fields in $\Phi_1$ against basic fields in $\Phi_2$. For each matching, we replace matched pairs of basic fields by their propagators (acted on by respective derivatives in $w,\bar{w},z,\bar{z}$ that were acting on those basic fields). We multiply the result with the unmatched basic fields (acted on by respective derivatives), while also replacing basic fields evaluated at $w$ with their Taylor expansion around $z$. Finally, we sum these contributions over all partial matchings. 

Using this recipe one obtains
the following OPEs of distinguished fields -- the stress-energy tensor $T=\ddel b\,\ddel c+a\,\ddel\lambda$, its BRST primitive $G=a\,\ddel b$ and the BRST current $J= \gamma\,\ddel c$ (and their anti-holomorphic counterparts $\bar{T},\bar{G},\bar{J}$):
\begin{eqnarray}
T(w)\, T(z) &\sim& \frac{2 T(z)}{(w-z)^2}+\frac{\ddel T(z)}{w-z}+\reg \label{TT OPE} \\
J(w)\, G(z) & \sim & -\frac{1}{(w-z)^3}+\frac{ (\gamma \, a)_z}{(w-z)^2}+\frac{T(z)}{w-z}+\reg  \label{JG OPE}\\
T(w)\, G(z) &\sim &  \frac{2 G(z)}{(w-z)^2}+\frac{\ddel G(z)}{w-z}+\reg \nonumber \\
T(w)\, J(z) &\sim  & \frac{J(z)}{(w-z)^2}+\frac{\ddel J(z)}{w-z}+\reg  \nonumber
\end{eqnarray}
and their complex conjugates.
Also, one has $G(w)\,G(z)\sim\reg$ and $J(w)\, J(z)\sim \reg$ 
All OPEs between holomorphic objects $T,G,J$ and anti-holomorphic objects $\bar{T},\bar{G},\bar{J}$ are regular. In particular, from the absence of $4$-th order pole in (\ref{TT OPE}), we see that the central charge of the theory vanishes $\s{c}=0$.\footnote{
Recall that a conformal field theory with central charge $\s{c}$ is characterized by the following OPE of the stress-energy tensor with itself: $T(w)\,T(z)\sim \frac{\s{c}/2}{(w-z)^4}+\frac{2T(z)}{(w-z)^2}+\frac{\ddel T(z)}{w-z}+\reg$ (plus the conjugate expression for $\bar{T}\bar{T}$, plus $T\bar{T}\sim\reg$). Also recall that a field $\Phi$ is \emph{primary}, of conformal weight $(h,\bar{h})$ iff its OPEs with the stress-energy tensor are: $T(w)\,\Phi(z)\sim \frac{h\,\Phi(z)}{(w-z)^2}+\frac{\ddel \Phi(z)}{w-z}+\reg$ and the conjugate
$\bar{T}(w)\,\Phi(z)\sim \frac{\bar{h}\,\Phi(z)}{(\bar{w}-\bar{z})^2}+\frac{\ddelb \Phi(z)}{\bar{w}-\bar{z}}+\reg$
} Also, we see that $G$ and $J$ are primary fields with conformal weights $(2,0)$ and $(1,0)$ respectively.

\begin{example}
For example, here is the computation of the OPE (\ref{JG OPE}). By (\ref{propagators}), we have the propagators $\lan \gamma_w\, a_z \ran=-\frac{1}{w-z} $ and $\lan (\ddel c)_w (\ddel b)_z\ran= \dd_w\dd_z \lan c_w\, b_z \ran =\dd_w\dd_z 2\log|w-z|=\frac{1}{(w-z)^2}$. In the OPE $( \gamma \dd c)_w (a \dd b)_z$ one gets three singular terms from the Wick contractions of either $\gamma_w$ with $a_z$ or $(\dd c)_w$ with $(\dd b)_z$ or both. Thus, 
\begin{multline}
( \gamma \dd c)_w (a \dd b)_z\sim \\ \sim
 \frac{-1}{w-z}\cdot \frac{1}{(w-z)^2}+\frac{1}{(w-z)^2} :\gamma_w a_z:+\frac{-1}{w-z} {:(\dd c)_w (\dd b)_z:} +\reg\sim\\
\sim
\frac{-1}{(w-z)^3}+\frac{ (\gamma a)_z}{(w-z)^2}+\frac{1}{w-z}( \dd\gamma\, a+\dd b\,\dd c )_z+\reg 
\end{multline} 
Here in the last step we replaced fields at $w$ with their Taylor expansions centered at $z$. Note that the products of fields at $w$ and  at $z$ occurring at the intermediate stage are normally ordered, i.e. Wick contractions inside them are prohibited. Finally, notice that $ \dd\gamma\, a+\dd b\,\dd c$ is equivalent to $T$ modulo equations of motion. Thus we obtain (\ref{JG OPE}).
\end{example}

For $\Phi(z)=\Phi(B,c)_z$ a $Q$-closed field (\ref{Q coh}), we obtain
\begin{equation*}
T(w) \Phi(z)\sim \frac{\dd \Phi(z)}{w-z}+\reg, \qquad \bar{T}(w) \Phi(z)\sim \frac{\ddelb \Phi(z)}{\bar{w}-\bar{z}}+\reg
\end{equation*}
Thus all polynomials in $B,c$ are primary fields of weight $(0,0)$.


\subsection{Extended Virasoro algebra}
Every 
field $\alpha$ which is holomorphic, i.e. satisfies $\ddelb\alpha\simeom 0$, and has conformal weight $(h,0)$,
determines \emph{mode operators}
\begin{equation}\label{operator associated to current}
\alpha^{(z)}_n:\quad \Phi(z) \mapsto \oint_{\mc{C}_z  
} \frac{dw}{2\pi i}\,(w-z)^{n+h-1}\,\alpha(w)\Phi(z)
\end{equation}
on the space of fields $\bb{F}_z$ where on the r.h.s. one has the integral in variable $w$ over a contour $\mc{C}_z$ going once counterclockwise around $z$.\footnote{The contour is supposed to be a boundary of a small neighborhood (e.g. a disk) of $z$, where ``small'' means that all the other field insertions in the correlators we are considering happen outside the neighborhood. Note that the 
holomorphic property
$\ddelb\alpha\simeom 0$ implies that one can deform the contour as long as it does not intersect with field insertions.}
In other words, $\alpha^{(z)}_n$ acts on a field $\Phi(z)$ by taking the coefficient of $(w-z)^{-n-h}$ 
in the OPE $\alpha(w)\,\Phi(z)$. I.e. one has the mode expansion -- the equality
\begin{equation}\label{mode expansion}
\alpha(w)=\sum_{n\in\ZZ+h} (w-z)^{-n-h}\alpha^{(z)}_n
\end{equation}
of $w$-dependent operators on $\bb{F}_z$. Here the left hand side acts on a field $\Phi(z)$ by sending it to the OPE $ \alpha(w)\Phi(z)$.
Similarly, 
an anti-holomorphic field $\bar\alpha$ (i.e. satisfying $\dd \bar\alpha\simeom 0$), of weight $(0,\bar{h})$,
determines mode operators
$\bar{\alpha}^{(z)}_n:\quad \Phi(z) \mapsto \oint_{\mc{C}_z } \frac{d\bar{w}}{-2\pi i}\,(\bar{w}-\bar{z})^{n+\bar{h}-1}\bar{\alpha}(w)\Phi(z)$.

An important case of this construction is for $\alpha$ a conserved (holomorphic) Noether current -- a primary field of conformal weight $(1,0)$ satisfying $\ddelb\alpha\simeom 0$. Then 
$$\hat\alpha^{(z)}:=\alpha^{(z)}_0:\quad \Phi(z)\mapsto \oint_{\mc{C}_z}\frac{dw}{2\pi i}\, \alpha(w) \Phi(z)$$ 
is the corresponding quantum 
Noether charge acting on $\bb{F}_z$.

In particular, it is a straightforward check that the operator $\hat{J}_\tot:=\hat{J}+\hat{\bar{J}}$ associated to the total BRST current\footnote{
Our normalization convention here is as follows: $\hat{J}_\tot\Phi(z):=-\frac{1}{4\pi}\oint_{\mc{C}_z} (J_\tot)_w \Phi_z=\oint_{\mc{C}_z}(\frac{dw}{2\pi i}J_w+\frac{d\bar{w}}{-2\pi i}\bar{J}_w)\Phi_z$. Here the factor $-\frac{1}{4\pi}$ is the same as the factor accompanying the action in the path integral (\ref{correlator}).
} 
$J_\tot=2i (dz \, J-d\bar{z}\, \bar{J})$ coincides with the classical BRST operator $Q
$ acting on $\bb{F}_z$.\footnote{
For example: $(\gamma\dd c)_w b_z\sim \frac{\gamma}{w-z}+\reg$ and $( \bar\gamma\ddelb c)_w b_z\sim \frac{\bar\gamma}{\bar{w}-\bar{z}}+\reg$, hence $\hat{J}b=\gamma, \hat{\bar{J}}b=\bar\gamma$, and thus $(\hat{J}+\hat{\bar{J}})b=\gamma+\bar\gamma=\lambda=Q(b)$. Likewise, $(\gamma\dd c)_w a_z\sim \frac{-\dd c}{w-z}+\reg$ and $( \bar\gamma\ddelb c)_w a_z\sim\reg$, hence $\hat{J}a=-\dd c,\hat{\bar{J}}a=0$ and thus $(\hat{J}+\hat{\bar{J}})a=-\dd c=Q(a)$. Another example is $\hat{J}_\tot G$ which is given by the residue in the OPE (\ref{JG OPE}), thus $\hat{J}_\tot G=T$ which is a confirmation that in the quantum setting the classical relation $T=Q(G)$ still holds.
}

One defines the Virasoro generators $L^{(z)}_n:={T}^{(z)}_n$ with $n\in \ZZ$, 
as the mode operators for the stress-energy tensor $T$, defined by (\ref{operator associated to current}) with $h=2$. Similarly, 
the anti-holomorphic Virasoro generators $\bar{L}^{(z)}_n$ are the mode operators for $\bar{T}$. 
We will also need the mode operators $G_n^{(z)}$ of the BRST-primitive $G$ (which also has weight $h=2$) and their conjugate counterparts $\bar{G}_n^{(z)}$ associated to $\bar{G}$.

\begin{example}\label{ex L_-1}
Operators $L^{(z)}_{-1}$ and $\bar{L}^{(z)}_{-1}$ act on $\bb{F}_z$ as partial derivatives: 
\begin{equation}\label{L_-1 = partial}
L^{(z)}_{-1}=\dd_z,\qquad \bar{L}^{(z)}_{-1}=\dd_{\bar{z}}
\end{equation}
\end{example}


From the OPEs between $T,G,J$ and their conjugates, one obtains the following super commutation relations (Lie brackets) for the graded Lie algebra linearly generated by the operators $Q,\{L_n\},\{G_n\},\{\bar{L}_n\},\{\bar{G}_n\}$:\footnote{We omit for brevity the superscript $(z)$, understanding that all operators here act on $\bb{F}_z$ for a fixed point $z\in\Sigma$.}
\begin{multline}\label{extended Virasoro relations}
[Q,Q]=0,\quad [L_n,L_m]=(n-m)L_{n+m}, \\ 
[Q,L_n]=0,\quad [Q,G_n]=L_n,\quad [L_n,G_m]= (n-m) G_{n+m},\quad [G_n,G_m]=0
\end{multline}
plus the conjugate relations. Commutators involving a holomorphic generator $\in\{L_n,G_n\}$ and an anti-holomorphic generator $\in\{\bar{L}_n,\bar{G}_n\}$ vanish. 
The degrees (ghost numbers) of the generators are:

\begin{center}
\begin{tabular}{|c|l|}
\hline
$Q$ & $+1$ \\
$L_n,\bar{L}_n$ & $0$ \\
$G_n,\bar{G}_n$ & $-1$ \\
\hline
\end{tabular}
\end{center}

In particular, this is an extension of the direct sum of two copies (coming from holomorphic and anti-holomorphic sectors) of Virasoro algebra with central charge $\s{c}=0$.

\begin{remark}
The theory contains a ``logarithmic field'' $(cb)_z$ whose OPE with $T$ is: $T(w)(cb)_z \sim \frac{1}{(w-z)^2}+\frac{\ddel (cb)_z}{w-z}+\reg$ Its presence implies that the Hamiltonian of the theory $\hat{H}=L_0+\bar{L}_0$ is not diagonalizable and has a Jordan block (with eigenvalue $0$) consisting of the eigenvector $1$ and a generalized eigenvector $\frac12 cb$.
\end{remark}

\begin{remark}[Sugawara construction] 
Consider the holomorphic 
fields $a,\gamma,\dd b,\dd c$ 
and consider their Fourier modes around $z$ defined by (\ref{mode expansion}).
Note that the stress-energy tensor $T= a\,\dd \gamma+\dd b\,\dd c$, BRST current $ J=\gamma\, \dd c$ and the primitive $G=a\,\dd b$ are explicitly written as  quadratic expressions in the four fields $a,\gamma,\dd b,\dd c$. Thus, for the Fourier modes we have 
\begin{multline}\label{Sugawara Vir via current}
L_n=\sum_m -(n-m) a_m \gamma_{n-m}+(\dd b)_m (\dd c)_{n-m}, \\
J_n=\sum_m \gamma_m (\dd c)_{n-m},\quad G_n=\sum_m a_m (\dd b)_{n-m}
\end{multline}
Commutation relations between the modes of $a,\gamma,\dd b,\dd c$ follow from OPEs $a_w\gamma_z\sim \frac{1}{w-z}+\reg$, $(\dd b)_w (\dd c)_z\sim \frac{-1}{(w-z)^2}+\reg$ between these fields. Explicitly, one has the commutation relations
\begin{equation}\label{Sugawara current}
[a_n,\gamma_m]=\delta_{n,-m},\quad [(\dd b)_n,(\dd c)_m]=m\,\delta_{n,-m}
\end{equation}
and all the other Lie brackets vanish. Formulae (\ref{Sugawara Vir via current}) can be seen as an analog of the Sugawara construction in Wess-Zumino-Witten theory, expressing Virasoro generators as quadratic combinations of  generators of the current algebra.
\end{remark}

\subsection{Witten's descent of observables}
We are interested in constructing ``$p$-observables'' -- composite fields $\OO^{(p)}$ with values in $p$-forms on $\Sigma$ with the property that 
\begin{equation}\label{descent eq}
Q\OO^{(p)}=d\OO^{(p-1)}
\end{equation}
for some $\OO^{(p-1)}$ and $d$ the de Rham operator. This would imply that, for $\gamma\subset \Sigma$ any $p$-cycle, the integral $\int_\gamma \OO^{(p)}$ is $Q$-closed; in particular, a correlator of several such expressions is a gauge-independent quantity.  Equation (\ref{descent eq}) is known as Witten's descent equation for observables \cite{Witten_revisited}.

One can solve equation (\ref{descent eq}) using operators $G_{-1},\bar{G}_{-1}$. Namely, we introduce the operator 
\begin{equation}\label{Gamma}
\Gamma=-dz\, G_{-1}-d\bar{z}\, \bar{G}_{-1}: \qquad \bb{F}_z\otimes \wedge^{p} T^*_z\Sigma\ra \bb{F}_z\otimes \wedge^{p+1} T^*_z\Sigma
\end{equation}
It can be viewed as the contraction of the de Rham operator $d=dz\,\ddel+d\bar{z}\,\ddelb$ with Fourier mode $-1$ of $G_\tot$.
By virtue of (\ref{extended Virasoro relations}) and (\ref{L_-1 = partial}), we have 
\begin{equation}\label{[Q,Gamma]}
[Q,\Gamma]=dz \,L_{-1}+d\bar{z}\,\bar{L}_{-1}=d 
\end{equation}
We fix a $Q$-closed $0$-observable $\OO^{(0)}\in\bb{O}_z$ -- a polynomial in $B$ and $c$, cf. (\ref{Q coh}), and construct 
\begin{eqnarray}\label{descent by Gamma}
\OO^{(1)}&:=&\Gamma\OO^{(0)}=-(dz\,G_{-1}+d\bar{z}\,\bar{G}_{-1})\,\OO^{(0)},\label{O^1}\\
\OO^{(2)}&:=& \frac12\, \Gamma^2 \OO^{(0)}=-dz\,d\bar{z} \,G_{-1}\,\bar{G}_{-1}\,\OO^{(0)} \label{O^2}
\end{eqnarray}
Observables $\OO^{(0)},\OO^{(1)},\OO^{(2)}$ satisfy the descent equation (\ref{descent eq}) for $p=0,1,2$.\footnote{
Indeed, the descent equation for $p=0$ reads $Q\OO^{(0)}=0$ which is satisfied by assumption. Next, for $p=1$, we have $Q\OO^{(1)}=[Q,\Gamma]\OO^{(0)}=d\OO^{(0)}$ by (\ref{[Q,Gamma]}). Finally, for $p=2$, we have $Q\OO^{(2)}=\frac12 Q\Gamma \OO^{(1)}=\frac12([Q,\Gamma]\OO^{(1)}+\Gamma Q \OO^{(1)})=\frac12 (d \OO^{(1)}+\Gamma d \OO^{(0)})=\frac12 (d\OO^{(1)}+d \OO^{(1)})=d\OO^{(1)}$. Here we used that $\Gamma$ commutes with $d=dz\,L_{-1}+d\bar{z}\, \bar{L}_{-1}$. 
}

Explicitly, $G_{-1}$ and $\bar{G}_{-1}$ act on $\bb{F}_z$ as derivations defined on generators by
\begin{align}
G_{-1}:\qquad c\mapsto -a,\;\gamma\mapsto  \dd b,
\quad \bar\gamma,b,a,\bar{a}\mapsto 0\\
\bar{G}_{-1}:\qquad c\mapsto -\bar{a},
\; \bar\gamma\mapsto  \ddelb b,\quad \gamma,b,a,\bar{a}\mapsto 0
\end{align}

Consider the case of an $N$-component theory (\ref{N-component BRST fiber}). For $\OO^{(0)}\in \bb{O}_z^{[N]}$ a polynomial in $B_1,\ldots,B_N,c^1,\ldots c^N$, the descended observables defined by (\ref{O^1},\ref{O^2}) are:
\begin{equation}\label{descent O^1}
\OO^{(1)}= \left((dz\,a^j+d\bar{z}\,\bar{a}^j)\frac{\dd}{\dd c^j}+(i\, dz\,\dd b_j-i\, d\bar{z}\, \ddelb b_j)\frac{\dd}{\dd B_j}\right)\,\OO^{(0)}=
\left(A^j\frac{\dd}{\dd c^j}-*db_j\,\frac{\dd}{\dd B_j}\right)\OO^{(0)}
\end{equation}
and
\begin{multline} \label{descent O^2}
\OO^{(2)}=-dz\,d\bar{z}\left(a^j\bar{a}^k\frac{\dd^2}{\dd c^j \dd c^k}+(ia^j\ddelb b_k+i\bar{a}^j \ddel b_k)\frac{\dd^2}{\dd c^j  \dd B_k} + \dd b_j \ddelb b_k\,\frac{\dd^2}{\dd B_j \dd B_k} \right)\OO^{(0)}\\
=\left(-\frac12 A^j\wedge A^k \frac{\dd^2}{\dd c^j \dd c^k}-A^j\wedge *db_k \frac{\dd^2}{\dd c^j \dd B_k}+ \frac12 *db_j\wedge *db_k\frac{\dd^2}{\dd B_j \dd B_k}\right)\OO^{(0)}
\end{multline}

\begin{example}\label{ex: Wilson loop} Taking $\OO^{(0)}=c$ (in $1$-component theory), we get $\OO^{(1)}=dz\,a+d\bar{z}\,\bar{a}=A$ and $\OO^{(2)}=0$. In particular, we can integrate this $1$-observable along  a closed oriented curve $\bm\gamma\subset \Sigma$, obtaining a $Q$-closed expression $\oint_{\bm\gamma} A$. Then one can, e.g., consider a correlator
$\lan B(z)\,\oint_{\bm\gamma} A \ran$.
The expression in the correlator is $Q$-closed and thus the correlator is topological -- invariant under isotopy.
Using the propagator $\lan B(z)A(w) \ran=2d_w \arg(w-z)$, we can compute this correlator:
\begin{equation}\label{linking number}
\lan B(z)\,\oint_{\bm\gamma} A \ran = 4\pi\, \mr{lk}(\bm\gamma,z)
\end{equation}
where $\mr{lk}(\bm\gamma,z)$ is the ``linking number'' -- the winding number of the curve $\bm\gamma$ around the point $z$.\footnote{Here we are implicitly assuming that $z$ is not on $\bm\gamma$. If $z\in \bm\gamma$, the correlator also makes sense: the linking number in (\ref{linking number}) then takes a half-integer value -- the half-sum of the values obtained by displacing $z$ normally to $\bm\gamma$ in two possible directions.}
\end{example}

\begin{example}
In $N$-component theory, consider 
\begin{equation*}
\OO^{(0)}=W(c)
\end{equation*} 
a polynomial in variables $c^j$ containing only monomials of even degree. Then we have
\begin{equation*}
\OO^{(1)}=A^j \frac{\dd}{\dd c^j}W(c),\qquad \OO^{(2)}=-\frac12 A^j\wedge A^k \frac{\dd^2}{\dd c^j \dd c^k}W(c)
\end{equation*}
This $2$-observable determines a deformation of the abelian theory analogous to the deformation of the Landau-Ginzburg model by a superpotential.
\end{example}

\begin{example}\label{ex: Bcc observable descent}
In $N$-component theory, consider the cubic observable 
\begin{equation*}
\OO^{(0)}=\frac12 f^i_{jk} B_i c^j c^k
\end{equation*}
with $f^i_{jk}$ arbitrary constant coefficients with $f^i_{jk}=-f^i_{kj}$. 
Note that, from the viewpoint of interpretation (\ref{observables as polyvectors}) of $0$-observables as polyvectors, this is a quadratic vector field on $\Pi \RR^N$.
Then we have
\begin{equation*}
\OO^{(1)}=f^i_{jk}B_i A^j c^k-\frac12 f^i_{jk}*db_i\,c^jc^k, \qquad \OO^{(2)}=\frac12 f^i_{jk}B_i A^j A^k -f^i_{jk}*db_i\,A^j c^k
\end{equation*}
This $2$-observable, in the case when $f^i_{jk}$ are the structure constants of a Lie algebra, determines the deformation of the abelian $BF$ theory into the non-abelian $BF$ theory.
\end{example}

\subsubsection{Descent vs. AKSZ construction}
Within Batalin-Vilkovisky formalism, the $N$-component abelian $BF$ theory is defined by the master action coming from the AKSZ construction \cite{AKSZ},
\begin{equation}\label{S_BV}
S^\mr{BV}=\int B_k dA^k + A^*_k dc^k + \lambda_k b^{*k} = \underbrace{\int \mc{B}_k d\mc{A}^k}_{S^\mr{AKSZ}} + \underbrace{\int \lambda_k b^{*k}}_{S^\mr{aux}}
\end{equation}
-- a function on the space of BV fields
\begin{equation}\label{F_BV}
\mc{F}^\mr{BV}=\mc{F}^\mr{AKSZ}\times\mc{F}^\mr{aux}=\mr{Map}(T[1]\Sigma,T^*[1]V[1])
\times \Big(\underbrace{\Omega^0}_{\lambda^k}\oplus \underbrace{\Omega^0[-1]}_{b_k}\oplus \underbrace{\Omega^2[-1]}_{\lambda^{*k}} \oplus \underbrace{\Omega^2}_{b^{*k}}\Big)
\end{equation}
Here $V=\RR^N$ is the coefficient space of the theory; $\Omega^p$ is a shorthand for $\Omega^p(\Sigma)\otimes V$. The first factor above -- the AKSZ mapping space is parameterized by the fields $c^k,A^k,B_k$ and the respective anti-fields $c^*_k,A^*_k,B^{*k}$ which assemble into two \emph{AKSZ superfields} -- nonhomogeneous forms on $\Sigma$,
\begin{equation*}
\mc{A}^k=c^k+A^k+B^{*k},\qquad \mc{B}_k=B_k+A^*_k+c^*_k
\end{equation*}
parameterizing the first and second term in $\mc{F}^\mr{AKSZ}=\Omega^\bullet(\Sigma,V[1])\oplus \Omega^\bullet(\Sigma,V^*)$.
Thus the entire field content in BV setting is:

\vspace{5pt}
\begin{tabular}{|l|ccc|ccc||cccc|}
\hline
field/antifield & $c^k$ & $A^k$ & $B^{*k}$ & $B_k$ & $A^*_k$ & $c^*_k$ & $\lambda_k$ & $b_k$ & $\lambda^{*k}$ & $b^{*k}$ \\
\hline
form degree on $\Sigma$ & 0 & 1 & 2 & 0 & 1 & 2 & 0 & 0 & 2 & 2 \\
ghost number & 1 & 0 & -1 & 0 & -1 & -2 & 0 & -1 & -1 & 0 \\
\hline
\end{tabular}
\vspace{5pt}

Here objects without stars are the BRST fields and objects with stars are the corresponding anti-fields. $\mc{F}^\mr{BV}$ carries the symplectic form of ghost number $-1$, ${\omega^\mr{BV}=\sum_\phi \int \delta\phi\wedge \delta\phi^*}$ where the sum is over all species of BRST fields, $\phi\in \{c^k,A^k,B_k,\lambda_k,b_k\}$. The action (\ref{S_BV}) satisfies the classical master equation 
\begin{equation*}
(S^\mr{BV},S^\mr{BV})=0
\end{equation*} 
with $(-,-)$ the Poisson bracket (the \emph{BV anti-bracket}) on functions on $\mc{F}^\mr{BV}$ associated to the symplectic structure $\omega^\mr{BV}$.

Imposing the Lorenz gauge-fixing corresponds in the BV language to restricting from the whole space of BV fields to a Lagrangian submanifold $\mc{L}\subset \mc{F}^\mr{BV}=T^*[-1]\mc{F}$ defined as the graph of the exact $1$-form $\delta \Psi$ on the space of BRST fields, with $\Psi$ the gauge-fixing fermion (\ref{Psi}). Explicitly, $\mc{L}$ is given by
\begin{equation}\label{L}
\mc{L}:\qquad \left\{ 
\begin{array}{l}
c^k,A^k,B_k,\lambda_k ,b_k \quad \mbox{are free} \\ 
A^*_k=-*db_k, \;\; b^{*k}=d*A^k,\;\; c^*_k=B^{*k}=\lambda^{*k}=0
\end{array}
\right.
\end{equation}
In particular the restriction $\left.S^\mr{BV}\right|_\mc{L}$ is exactly the gauge-fixed action (\ref{S}).

Denote by $\mc{X}=T^*[1]V[1]=V[1]\oplus V^*$ the target of the AKSZ mapping space, appearing in (\ref{F_BV}). Let $\mr{ev}:\mc{F}^\mr{AKSZ}\times T[1]\Sigma \ra \mc{X}$ be the evaluation map for the AKSZ mapping space. Looking at (\ref{observables as polyvectors}) and our computation of the descent (\ref{descent O^1},\ref{descent O^2}), we make the following observations:
\begin{enumerate}[(i)]
\item The space of $0$-observables  $\bb{O}_z$ (\ref{observables as polyvectors}) coincides with the space of functions on the AKSZ target $\mc{X}$. 
\item For any $0$-observable, $\OO^{(0)}\in \bb{O}_z$, adding to it its first and second descent, we obtain the pullback of $\OO^{(0)}$, regarded as a function on the AKSZ target, by the evaluation map:
\begin{equation}\label{O^0+O^1+O^2= AKSZ pullback}
\OO^{(0)}+\OO^{(1)}+\OO^{(2)} =\left. \mr{ev}^*\OO^{(0)} \right|_\mc{L}
\end{equation}
For example, for $\OO^{(0)}=c^k$, (\ref{O^0+O^1+O^2= AKSZ pullback}) yields $c^k+A^k=\mc{A}^k|_\mc{L}$ -- the restriction of the AKSZ superfield to the Lagrangian (\ref{L}). Likewise, for $\OO^{(0)}=B_k$, we get $B_k-*db_k = \mc{B}_k|_\mc{L}$.
\item As immediately implied by the previous point, a deformation $S\ra S+g\int \OO^{(2)}$ of the abelian $BF$ action by a $2$-observable is the same as turning on the target Hamiltonian 
$g\OO^{(0)}$ in AKSZ construction, i.e., adding to the BV action the term $g\int \mr{ev}^*\OO^{(0)}$ and imposing the Lorenz gauge by restricting to the Lagrangian (\ref{L}).
\end{enumerate}

\begin{remark}\label{rem: G_xi}
To any vector field $\xi$ on $\Sigma$ one can associate the following function of BV fields of ghost number $-2$:
\begin{equation*}
\bb{G}_\xi:=\int c^*_k\, \iota_\xi A^k + A^*_k\, \iota_\xi B^{*k} + db_k\, \iota_\xi \lambda^{*k}
\end{equation*}
The object $\bb{G}_\xi$ is the generator 
of the action of the vector field $\xi$, regarded as an infinitesimal diffeomorphism of $\Sigma$, on BV fields as a BV gauge transformation. I.e., one has
\begin{equation*}
((S^\mr{BV},\bb{G}_\xi),-) = \sum_\psi \int L_\xi \psi\, \frac{\delta}{\delta \psi}
\end{equation*}
-- the lifting of the Lie derivative $L_\xi$ operating on the BV fields to a vector field on $\mc{F}^\mr{BV}$; here $\psi$ runs over all species of BV fields.
This is an adaptation of the construction of \cite{Getzler_spinning} to the model in question.
One has the following relation between $\bb{G}_\xi$ and the descent operator $\Gamma$ (\ref{Gamma}):
\begin{equation}\label{G_xi vs Gamma}
(\bb{G}_\xi,\phi_z)\Big|_\mc{L}\;\; 
=
\;\; (\iota_\xi\, \Gamma)\circ \phi_z
\end{equation}
for $\phi\in \{A^k,B_k,c^k,\lambda_k, b_k\}$ any BRST field.
\end{remark}

\textbf{Reformulation with auxiliary fields extended to AKSZ superfields.}
Note that in (\ref{S_BV},\ref{F_BV}) the BV system is presented as a sum of an AKSZ system and an auxiliary system
which is not of AKSZ form. One can in fact cast the auxiliary system in AKSZ form, too, by extending the four auxiliary fields $b_k,\lambda_k,b^{*k},\lambda^{*k}$ to a quadruple of AKSZ superfields:
%
\begin{eqnarray*}
\wh{\lambda_k} &=& \boxed{\lambda_k}+\mu_k+\nu_k\\ 
{\wh {\lambda^{*k}}} &=& \nu^{*k}+\mu^{*k}+\boxed{\lambda^{*k}}\\
\wh{b_k} &=& \boxed{b_k}+f^*_k+e^*_k\\ 
\wh{b^{*k}} &=& e^k+f^k+\boxed{b^{*k}}
\end{eqnarray*}
The form degrees and ghost numbers of the field components here are as follows.

\vspace{5pt}
\begin{tabular}{|l|ccc|ccc|ccc|ccc|}
\hline
field/antifield & $\lambda_k$ & $\mu_k$ & $\nu_k$ & $\nu^{*k}$ & $\mu^{*k}$ & $\lambda^{*k}$ & $b_k$ & $f^*_k$ & $e^*_k$ & $e^k$ & $f^k$ & $b^*_k$ \\
\hline 
form degree on $\Sigma$ & 0 & 1 & 2 & 0 & 1 & 2 & 0 & 1 & 2 & 0 & 1 & 2 \\
ghost number & 0 & -1 & -2 & 1 & 0 & -1 & -1 & -2 & -3 & 2 & 1 & 0 \\ \hline 
\end{tabular}
\vspace{5pt}

The BV action (\ref{S_BV}) in this setting is replaced with the full AKSZ action\footnote{
In fact, one can write a simpler action $\bar{S}=\int \mc{B}_k\, d\mc{A}^k + \wh{\lambda_k}\, \wh{b^{*k}}$, which is BV canonically equivalent to $\til{S}$ by $\til{S}=\bar{S}+\Big(\bar{S} , \int \wh{\lambda^{*k}} \, d\wh{b_k}\Big)$.
}
\begin{equation}\label{S_AKSZfull}
\til{S}=\int \mc{B}_k\, d\mc{A}^k + \wh{\lambda^{*k}} \, d \wh{\lambda_k}+\wh{b^{*k}}\, d \wh{b_k}+\wh{\lambda_k}\, \wh{b^{*k}}
\end{equation}
-- a function on the full AKSZ mapping space 
\begin{equation*}
\til{\mathcal{F}}=\mr{Map}(T[1]\Sigma \;,\; \underbrace{T^*[1]V[1]}_{\mc{X}}\times \underbrace{T^*[1](V^*[-1]\oplus V^*)}_{\mc{X}^\mr{aux}})
\end{equation*}
The target $\mc{X}^\mr{full}=\mc{X}\times \mc{X}^\mr{aux}=T^*[1](V[1]\oplus V^*[-1]\oplus V^*)$ is a shifted cotangent bundle with base coordinates $\underline{c}^k, \underline{b}_k,\underline{\lambda}_k$ (corresponding to the superfields $\mc{A}^k$, $\wh{b_k}$, $\wh{\lambda_k}$) and fiber coordinates $\underline{B}_k,\underline{e}^k,\underline{\nu}^{*k}$ (corresponding to superfields $\mc{B}_k$, $\wh{b^{*k}}$, $\wh{\lambda^{*k}}$). Kinetic term of (\ref{S_AKSZfull}) corresponds to the standard canonical $1$-form on the target (as a cotangent bundle); term $\wh{\lambda_k}\, \wh{b^{*k}}$ corresponds to the target Hamiltonian $\Theta=\underline{\lambda}_k\,\underline{e}^k$.\footnote{
Note that passing to the cohomology of the cohomological vector field $Q_\mr{target}=(\Theta,-)$ acting on functions on $\mc{X}^\mr{full}$ contracts the auxiliary part of the target and yields functions on $\mc{X}$, or the space of $0$-observables $\bb{O}_z$.}
The gauge-fixing Lagrangian is $\til{\mathcal{L}}=\mr{graph}(\delta\Psi)$ in $\til{\mathcal{F}}$, regarded as the cotangent bundle to the space of non-starred fields, with $\Psi$ as before (\ref{Psi}) (viewed as a constant function in fields 
$\mu_k, \nu_k, e^k, f^k$):
\begin{equation*}
\til{\mc{L}}:\qquad \left\{
\begin{array}{l}
c^k,A^k,B_k,\lambda_k,\mu_k,\nu_k,b_k,e^k,f^k\quad \mbox{are free}, \\
A^*_k=-*db_k, \;\; b^{*k}=d*A^k,\\ c^*_k=B^{*k}=\lambda^{*k}=\mu^{*k}=\nu^{*k}=f^*_k=e^*_k=0
\end{array}
\right.
\end{equation*}
Restriction of the action (\ref{S_AKSZfull}) to $\til{\mc{L}}$ yields
\begin{equation*}
\left. \til{S}\right|_{\til{\mc{L}}} = S+ \int \Big((\mu_k+db_k) f^k + \nu_k e^k\Big)
\end{equation*}
with $S$ the BRST action (\ref{S}). Integrating out the fields $\mu_k,\nu_k,e^k,f^k$, we obtain the action $S$.

In this setting for abelian $BF$ theory, with auxiliary AKSZ superfields, the generator of an infinitesimal diffeomeorphism (cf. Remark \ref{rem: G_xi}) is:
\begin{multline*}
\til{\bb{G}}_\xi=\int \mc{B}_k\; \iota_\xi \mc{A}^k+\wh{\lambda^{*k}}\;\iota_\xi \wh{\lambda_k}+
\wh{b^{*k}}\,\iota_\xi \wh{b_k} \\
=\int c^{*}_k\,\iota_\xi A^k + A^*_k \,\iota_\xi B^{*k}+\lambda^{*k}\,\iota_\xi \mu_k + 
\mu^{*k}\,\iota_\xi \nu_k + b^{*k} \,\iota_\xi f^*_k + f^k\,\iota_\xi e^*_k
\end{multline*}
Relation (\ref{G_xi vs Gamma}) holds again in this setting, modulo equations of motion.

\subsubsection{Towards Gromov-Witten invariants: a toy example}
Correlators of the form 
\begin{equation*}
\big\langle G_\tot(w_1)\cdots G_\tot(w_p)\; \Phi_1(z_1)\cdots \Phi_n(z_n) \big\rangle
\end{equation*}
with $Q$-closed fields $\Phi_1,\ldots,\Phi_n$ define closed $p$-forms on $\mc{M}_{\Sigma,n}$ -- 
the moduli space of conformal structures on $\Sigma$ with $n$ marked points $z_1,\ldots,z_n$ (here $\Sigma$ can be any surface). Integrating such a form over a $p$-cycle on $\mc{M}_{\Sigma,n}$, one obtains interesting periods -- a version of Gromov-Witten invariants. Example \ref{ex: Wilson loop} above leads to a simple example of such a period.

\begin{example}\footnote{
This example is a corrected version of the one given in the previous version of the paper. The old example contained a mistake: the descent was applied only to one field in the correlator, which led to a $1$-form on the configuration space which was $PSL_2(\CC)$-invariant but not horizontal, and hence did not descend to the moduli space.
}
For $\Sigma=\CC P^1$, consider the correlator 
\begin{equation}\label{GW_Xi}
\rho=\langle\; \Gamma \big(c(z_0) B(z_1) \til{\Theta}(z_2) c(z_3)\big)\; \rangle_{\CC P^1}
\end{equation}
where we understand that the descent operator $\Gamma$ (\ref{Gamma}) acts on a product of fields as a derivation. This is a correlator on the sphere rather than on a plane, with $\til\Theta(z_2)=\delta(b)\delta(\gamma)\delta(\bar\gamma)\big|_{z_2}$ and $c(z_3)=\delta(c)|_{z_3}$ the ``soaking operators'' for zero-modes of the kinetic operators 
$\del,\delb, \del\delb$ in the action. We refer the reader to section 2.4 in \cite{LMY2} for further details, and to section 10 of \cite{Witten_12} for the general technology of 
correlators involving delta-functions of fields
 in $\beta\gamma$-systems. Explicit evaluation of $\rho$ yields
\begin{equation*}
\rho=2\, d\ \mr{arg} \frac{(z_0-z_1)(z_2-z_3)}{(z_0-z_2)(z_1-z_3)}\qquad \in \Omega^1(\mr{Conf}_{4}(\CC P^1))
\end{equation*}
Note that here the four points on $\CC P^1$ enter via their cross-ratio. In particular, $\rho$ is a closed \emph{$PSL_2(\CC)$-basic} $1$-form on the open configuration space of $4$ points on $\CC P^1$ and thus descends to a closed $1$-form on the (non-compactified) moduli space $\mc{M}_{0,4}$ of conformal structures on $\CC P^1$ with $4$ marked points. 

As an example of a period: integrating $\rho$  over $z_0$ in a contour $\bm\gamma \subset \CC P^1-\{z_1,z_2,z_3\}$, we get the ``linking number'' of $\bm\gamma$ and the $0$-cycle $[z_1]-[z_2]$ -- the difference of winding numbers of $\bm\gamma$ around $z_1$ and around $z_2$:\footnote{
Note that the winding number of a closed curve $\bm\gamma$ around a point $z_k$ is well-defined on a plane but not on a sphere. However, the difference is well-defined on a sphere. I.e., mapping the sphere onto the plane by stereographic projection with ``North pole'' at $p\in \CC P^1 -\bm\gamma\cup \{z_1,z_2,z_3\}$, we prescribe values to winding numbers which jump when $p$ crosses $\bm\gamma$, but the difference of winding numbers does not jump. More generally, on the sphere one has a well-defined linking number $\mr{lk}(\bm\gamma,\xi)$ of a closed curve and a $0$-cycle $\xi=\sum_{k=1}^n\alpha_k [z_k]$ if and only if the sum of coefficients vanishes, $\sum_k \alpha_k=0$.
}
\begin{equation*}
\oint_{\bm\gamma\ni z_0}\rho = 4\pi 
\ \mr{lk}(\bm\gamma,[z_1]-[z_2])
\end{equation*}
One can see this computation as an integral of a closed $1$-form on $\mc{M}_{0,4}$ over  a $1$-cycle 
$\bm\gamma$ 
in the fiber of the projection $\mc{M}_{0,4}\ra \mc{M}_{0,3}=\ast$ forgetting the point $z_0$, yielding the linking number as a simplest Gromov-Witten period.

As a small variation on this example, instead of a single insertion of $B$ in (\ref{GW_Xi}), we can consider several insertions of $e^{\alpha B}$, with $\alpha$ some coupling constants, which gives the period
$$
\int_{\bm\gamma\ni z_0} \Big\langle\; \Gamma \Big( c(z_0) e^{\alpha_1 B(z_1)}\cdots e^{\alpha_n B(z_n)} \til\Theta(z_{n+1}) c(z_{n+2}) \Big)\; \Big\rangle = 4\pi
\ \mr{lk}\Big(\bm\gamma, \sum_{k=1}^n\alpha_k [z_k]- \big(\sum_{k=1}^n\alpha_k\big) [z_{n+1}]\Big)
$$
Here on the l.h.s. we integrate a closed $1$-form on $\mc{M}_{0,n+3}$ (corresponding to the $PSL_2(\CC)$-basic $1$-form $2\sum_{k=1}^n \alpha_k d\ \mr{arg} \frac{(z_0-z_k)(z_{n+1}-z_{n+2})}{(z_0-z_{n+1})(z_k-z_{n+2})} $ on the open configuration space of $n+3$ points on $\CC P^1$) over a $1$-cycle $\bm\gamma$ in the fiber of $\mc{M}_{0,n+3}\ra \mc{M}_{0,n+2}$.
\end{example}

We plan to return to a detailed discussion of Gromov-Witten invariants in the context of abelian and non-abelian $BF$ theory in a future paper.

\subsection{BV algebra structure on the space of $0$-observables 
}\footnote{We refer the reader to \cite{Getzler_BV} for the details on emergence of the BV structure in the context of twisted superconformal theory.}
The space of $0$-observables $\bb{O}_z$ in addition to being  
a graded commutative algebra has a degree $-1$ Poisson bracket defined by
\begin{equation} \label{Poisson bracket}
\{\OO_1,\OO_2\}:= \frac{(-1)^{|\OO_1|}}{4\pi}\oint_{\mc{C}_z} (\Gamma\OO_1)_w (\OO_2)_z
\end{equation}
I.e. one descends the first $0$-observable to a $1$-observable and integrates over a contour encircling the second $0$-observable.
\begin{example}
For $\OO_1=c$, $\OO_2=B$, we have $\Gamma(c)=A$ and we obtain 
$$\{c,B\}=-\frac{1}{4\pi} \oint_{\mc{C}_z}\underbrace{A_w B_z}_{\sim 2d_w\arg(w-z)+\reg}=-1$$
cf. example \ref{ex: Wilson loop}. In particular, $c$ and $B$ are conjugate variables for the Poisson bracket.
\end{example}
Explicitly, the Poisson bracket (\ref{Poisson bracket}) is:
\begin{equation*}
\{\OO_1,\OO_2\}=\frac{\dd}{\dd B}\OO_1\; \frac{\dd}{\dd c} \OO_2+(-1)^{|\OO_1|}\frac{\dd}{\dd c} \OO_1\; \frac{\dd}{\dd B}\OO_2
\end{equation*}
Commutative multiplication together with this bracket comprise the structure of a ``$P_2$ algebra'' on $\bb{O}_z$ (the algebra over the homology of the operad $E_2$ of little $2$-disks\footnote{
Recall that $E_2$ is a topological operad with the space of $n$-ary operations $E_2(n)$ being the space of configurations $\mb{o}$ of $n$ ordered disjoint disks inside the unit disk in $\RR^2\simeq \CC$; an $i$-th composition $\mb{o}_1\circ_i \mb{o}_2$ of two operations corresponds to fitting a rescaled disk configuration $\mb{o}_2$ instead of $i$-th disk in $\mb{o}_1$. It is instructive to think of a disk configuration $\bf{o}$, considered modulo rescalings, as a $2$-dimensional cobordism from $n$ in-circles (boundaries of the inner disks) to the single out-circle, with composition in $E_2$ corresponding to composition (gluing) of cobordisms.  The \emph{framed} operad of little $2$-disks, $E_2^\mr{fr}$ is defined in the same way where additionally one marks a point on the boundary of each disk, and the outer unit disk comes with the standard marked point $1\in\CC$. Then in the composition, in addition to rescaling a disk configuration, one does a rotation so as to fit the marked point on the out-disk of $\bf{o}_2$ with the marked point on the $i$-th disk of $\bf{o}_1$.
}). 

In addition to the bracket $\{-,-\}$, one has the operator 
\begin{equation}\label{G_0^-}
G_0^-:=\frac{1}{2i}(G_0-\bar{G}_0)
\end{equation}
-- the contraction of $G_\tot$ with the vector field corresponding to rotation about the point $z$, 
acting on $\bb{F}_z$ and in particular on $\bb{O}_z$.
\begin{example}
E.g. acting on $\OO=Bc$, one has 
\begin{multline*}
G_w \OO_z=(a\dd b)_w(Bc)_z\sim \\ 
\sim \frac{-i}{w-z}\,\frac{-1}{w-z}+\frac{-i}{w-z}(\dd b)_w\, c_z+\frac{-1}{w-z}a_w B_z+\reg 
\sim \quad \frac{i}{(w-z)^2}+\frac{(-i\dd b\,c-aB)_z}{w-z}+\reg
\end{multline*}
Thus, $G_0(Bc)=i$ -- the coefficient of the second order pole in the OPE above, and similarly one obtains $\bar{G}_0(Bc)=-i$. Therefore, $G^-_0(Bc)=1$.
\end{example}
Explicitly,
the operator $G_0^-$ acts on $0$-observables by
\begin{equation*}
G_0^-:\quad \OO\mapsto \frac{\dd^2}{\dd B\, \dd c}\OO
\end{equation*}
Thus, one recognizes in $G_0^-$ the Batalin-Vilkovisky Laplacian (of degree $-1$) and hence $(\bb{O}_z,\cdot,\{-,-\},\Delta)$ is a BV algebra with the bracket and the Laplacian of degree $-1$. In other words, it is an algebra over the homology of the operad $E_2^\mr{fr}$ of \emph{framed} little $2$-disks.

Note that, from the standpoint of identification of $0$-observables with polyvectors (\ref{observables as polyvectors}), this is the standard BV algebra structure on polyvectors.

\begin{remark}
In the example \ref{ex: Bcc observable descent}, we expect $\OO^{(2)}$ to give a classically consistent deformation of the action $S\mapsto S+g\int_\Sigma \OO^{(2)}$ (with $g$ the deformation parameter) if and only if $f^i_{jk}$ satisfy Jacobi identity, i.e., define a Lie algebra on the space of coefficients $\RR^N$ and we expect the deformation to be consistent on the quantum level if additionally the unimodularity property $f^i_{ij}=0$ holds. Note that these two cases correspond to, respectively, classical and quantum BV master equation holding for $\OO^{(0)}=\frac12 f^i_{jk}B_i c^j c^k$:
\begin{equation*}
\{\OO^{(0)},\OO^{(0)}\}=0,\qquad 
G_0^- \OO^{(0)}=0
\end{equation*}
\end{remark}

\begin{remark}
One can consider $S^1$-equivariant version of BRST cohomology (\ref{Q coh}) -- cohomology of the equivariant extension of the BRST operator
\begin{equation*}
Q_{S^1}:=Q+\epsilon G^-_0
\end{equation*}
acting on the kernel of $L_0^-\propto Q_{S^1}^2$ in $\bb{F}_z[\epsilon]$ (rotationally-invariant fields valued in polynomials in $\epsilon$), with $\epsilon$ the degree $2$ equivariant parameter. This equivariant cohomology evaluates, in the context of $N$-component theory, to
\begin{equation*}
H_{S^1}(\bb{F}_z)\simeq H_{\epsilon G^-_0}(\bb{O}_z[\epsilon])=T_\mr{poly}^\mr{div-free}(\Pi\RR^N)\oplus c^1\ldots c^N\cdot \epsilon \CC[\epsilon]
\end{equation*}
-- the space of divergence-free (or ``unimodular'') polyvectors on $\Pi \RR^N$,
plus the $\CC[\epsilon]$-linear span of the products of all ghosts times $\epsilon$.
\end{remark}

\subsubsection{Structure of an algebra over the framed $E_2$ 
operad on the space of composite fields}
The space of composite fields itself $\bb{F}_z$ has the structure of an algebra over the operad $E_2^\mr{fr}$ of framed little $2$-disks. Namely, given a configuration $\mb{o}\in E_2^\mr{fr}(n)$ of $n$ framed disks with centers at $z_1,\ldots,z_n$, radii $r_1,\ldots,r_n$ and rotation angles $\theta_1,\ldots,\theta_n$, one constructs a map 
\begin{equation}\label{E_2 operation}
\mb{o}: \bb{F}_{z_1}\otimes \cdots \otimes \bb{F}_{z_n} \ra \bb{F}_0
\end{equation}
which sends an $n$-tuple of composite fields $\Phi_1(z_1),\ldots,\Phi_n(z_n)$ to a field $\Psi\in \bb{F}_0$ characterized by the property that 
\begin{equation}\label{E_2 op definition}
\lan \left(\prod_{j=1}^n r_j^{\hat H^{(z_j)}}e^{i\theta_j \hat P^{(z_j)}}\Phi_j(z_j) \right) \cdot \phi_1(y_1)\cdots \phi_m(y_m) \ran=
\lan \Psi(0) \cdot \phi_1(y_1)\cdots \phi_m(y_m) \ran
\end{equation}
for any test fields $\phi_1,\ldots,\phi_m$ inserted at points $y_1,\ldots,y_m$ outside the unit disk on $\CC$. Here $\hat{H}^{(z)}:=L_0^{(z)}+\bar{L}_0^{(z)}$ and $\hat{P}^{(z)}:=L_0^{(z)}-\bar{L}_0^{(z)}$ are the energy and momentum operators acting on fields at $z$; in particular, for a field $\Phi(z)$ of conformal weights $(h,\bar{h})$, the rescaling factor in the l.h.s. of (\ref{E_2 op definition}) is 
\begin{equation}\label{E_2 scaling}
r^{h+\bar{h}}e^{i\theta (h-\bar{h})}
\end{equation} 
Thus, operation (\ref{E_2 operation}) is an $n$-point version of an operator product expansion, rescaled appropriately to account for the size and orientation of the disks.

One calculates operations (\ref{E_2 operation}) explicitly using Wick's lemma: one considers all partial contractions between basic fields in $\Phi_1,\ldots,\Phi_n$, replaces those with the appropriate propagators and replaces all the remaining fields with their Taylor expansion at zero.
Finally, one rescales the result with the factors (\ref{E_2 scaling}) (we are assuming for simplicity that fields $\Phi_j$, with $1\leq j\leq n$, have well-defined conformal weights, i.e., are eigenvectors for the operators $L_0,\bar{L}_0$).

\begin{example} Let $\mb{o}$ be a configuration of two disks centered at $z_1,z_2$ with radii $r_1,r_2$ and rotation angles $\theta_1,\theta_2$, and let $\Phi_1=J=\gamma\,\dd c$ and $\Phi_2=G=a\,\dd b$. Recall that the conformal weights are $(h,\bar{h})=(1,0)$ for $J$ and $(h,\bar{h})=(2,0)$ for $G$. We obtain from Wick's lemma
\begin{multline*}
\mb{o}(J\otimes G)=\\
=r_1e^{i\theta_1}(r_2 e^{i\theta_2})^2\left(
-\frac{1}{(z_1-z_2)^3}+\frac{:\gamma_{z_1}a_{z_2}:}{(z_1-z_2)^2}-\frac{:(\dd c)_{z_1}(\dd b)_{z_2}:}{z_1-z_2}+:(\gamma\dd c)_{z_1}(a\dd b)_{z_2}:
\right) \\
=r_1e^{i\theta_1}(r_2 e^{i\theta_2})^2\left(
-\frac{1}{(z_1-z_2)^3}+\sum_{k,l\geq 0}\frac{z_1^k z_2^l}{k!\,l!}\left(\frac{\dd^k\gamma\, \dd^l a}{(z_1-z_2)^2}-\frac{\dd^{k+1} c\, \dd^{l+1} b}{z_1-z_2}+\dd^k(\gamma\dd c)\,\dd^l(a\dd b)
\right)
\right)
\end{multline*}
Here in the last expression all fields are evaluated at $z=0$. Note that the Taylor series in $k,l$ converges under the correlator with test fields inserted at points outside the unit disk, using that $z_1,z_2$ are inside the unit disk.
\end{example}

This way one constructs the $E_2^\mr{fr}$-algebra structure on the space of composite fields $\bb{F}_z$. Extending it by linearity, one gets the action of singular $0$-chains of $E_2^\mr{fr}$ on $\bb{F}_z$. In a similar way one constructs the action of all chains $C_\bullet(E_2^\mr{fr})$ on $\bb{F}_z\otimes \wedge^\bullet T^*_z\Sigma$ -- composite fields with values in differential forms: one constructs the following differential form on $E_2^\mr{fr}(n)$ with values in products of fields:
\begin{equation}\label{E_2 form}
\prod_{j=1}^n \underbrace{\zeta_j^{L_0}\left(1-\frac{d\zeta_j}{\zeta_j} G_0\right) 
\zeta_j^{\bar{L}_0}\left(1-\frac{d\bar\zeta_j}{\bar\zeta_j} \bar{G}_0\right)}_{=\exp{\left[Q-d_\zeta\, ,\, G_0\log\zeta_j+\bar{G}_0 \log\bar\zeta_j\right]}} {\bm\Phi}_j(z_j)
\end{equation}
and integrates it over the chain in $E_2^\mr{fr}$. This construction is considered under the correlator with an arbitrary collection of test fields outside the unit disk, as in (\ref{E_2 op definition}).
Here $\zeta_j=r_je^{i\theta_j}$ and $\bm\Phi_j(z_j)$ are composite fields with values in differential forms on $\Sigma$; we suppressed the superscripts $(z_j)$ for the operators $L_0,G_0$ and their conjugates. 

Further, one can restrict the construction above (\ref{E_2 form}) to fields of form 
\begin{equation}\label{Phi extended by descents}
\bm\Phi(z)=\Phi(z)+\Gamma \Phi(z)+\frac12 \Gamma^2 \Phi(z)=e^\Gamma \Phi(z)
\end{equation} 
with $\Gamma$ as in (\ref{Gamma}) -- i.e. sums of an ordinary (not form-valued) composite field and its first and second descents. This way we obtain a representation of $E_2^\mr{fr}$ as a \emph{differential graded} operad on the space of composite fields $\bb{F}_z$ (not form-valued), viewed as a cochain complex with BRST differential $Q$.\footnote{
Indeed, denote the form (\ref{E_2 form}) evaluated on fields of form (\ref{Phi extended by descents}) by $\Xi(\Phi_1,\ldots,\Phi_n)$. Then we have, by construction, $(d-Q)\Xi(\Phi_1,\ldots,\Phi_n)=\sum_{j=1}^n\pm\,\Xi(\Phi_1,\ldots, Q\Phi_j,\ldots,\Phi_n)$. With $d=\sum_j d_{\zeta_j}+d_{z_j}$ the de Rham differential on $E_2^\mr{fr}$ and $\pm$ the Koszul signs. Therefore, for $C\subset E_2^\mr{fr}$ a chain, one has $\int_{\dd C}\Xi(\Phi_1,\ldots,\Phi_n)=\int_C d\Xi=\int_C \big( Q\Xi(\Phi_1,\ldots,\Phi_n)-\sum_j\pm\,\Xi(\Phi_1,\ldots,Q\Phi_j,\ldots,\Phi_n)\big)$. Thus the map 
$C_\bullet(E_2^\mr{fr}(n))\mapsto \mr{Hom}(\bb{F}^{\otimes n},\bb{F})$ 
sending a chain $C$ to the  multilinear map $\Phi_1\otimes\cdots\otimes \Phi_n\mapsto \int_C \Xi (\Phi_1,\ldots,\Phi_n)$ is a chain map.
}

Passing to (co)homology, we get the action of the homology $H_\bullet(E_2^\mr{fr})$ on $H^\bullet_Q(\bb{F}_z)=\bb{O}_z$ -- the BV algebra structure $(\bb{O}_z,\cdot,\{-,-\},G_0^-)$ described above.

\begin{remark} In the discussion of the $E_2^\mr{fr}$-action on composite fields and BV algebra structure on $0$-observables, we used only a part of the extended Virasoro symmetry of the space of composite fields -- only modes $n=-1$ and $n=0$ (which displace and rotate/dilate the disks). Using the rest of the modes, one can infinitesimally 
reparameterize and deform the disks and thus, integrating the Virasoro action, one can
extend the $E_2^\mr{fr}$-action to the action of (the chains of) a larger operad of general genus zero conformal cobordisms 
$
S^1\sqcup\cdots\sqcup S^1\ra S^1$ with parameterized boundaries (more precisely, the operad of Riemannian spheres with $n+1$ disjoint conformally embedded disks -- Segal's genus zero operad).
In particular, in Segal's picture of conformal field theory \cite{Segal}, the complex $(\bb{F},Q)$ is the non-reduced space of states associated to a circle and $\bb{O}$ is the reduced space of states.
\end{remark}

\begin{remark}
Note that the normally-ordered version of the expression (\ref{E_2 form}) evaluated on fields (\ref{Phi extended by descents}) can be rewritten as follows:
\begin{equation*}
:\Xi(\Phi_1,\ldots,\Phi_n):=\prod_{j=1}^n 
\exp\Big[Q-d\, ,\, \log\zeta_j\,G_0+z_j\,G_{-1}+ \log\bar\zeta_j\,\bar G_0+\bar z_j\,\bar G_{-1}\Big]
\quad\circ \Phi_j(0)
\end{equation*}
where $d$ is the de Rham operator on $E_2^\mr{fr}$, i.e. the total de Rham operator in variables $z_j,\zeta_j$ (and conjugates).
\end{remark}

\subsection{The $U(1)$-current and twisting back to a superconformal field theory} \label{sec: U(1) current and untwist}
Consider the field 
\begin{equation}
j= \gamma a
\end{equation}
which we have encountered as a coefficient of the second order pole in the $J(w)G(z)$ OPE (\ref{JG OPE}). It is the Noether current for the $U(1)$-symmetry of the action, which rotates the phases of the fields $a,\gamma$ in the opposite directions: $a\mapsto e^{i\theta}a$, $\gamma\mapsto e^{-i\theta}\gamma$ and does not touch the  fields $b,c,\bar{a},\bar\gamma$.
This current is conserved modulo equations of motion, $\ddelb j\simeom 0$,
and satisfies the following OPEs:
\begin{eqnarray}
T_w j_z &\sim& \frac{1}{(w-z)^3}+\frac{j_z}{(w-z)^2}+\frac{\dd j_z}{w-z}+\reg \label{Tj OPE}\\ 
j_w J_z& \sim& \frac{J_z}{w-z}+\reg \nonumber \\ 
j_w G_z&\sim& \frac{-G_z}{w-z}+\reg \nonumber \\ 
j_w j_z&\sim& \frac{-1}{(w-z)^2}+\reg \label{jj OPE}
\end{eqnarray}
In particular, 
the fields $J$ and $G$ have charges $+1$ and $-1$ respectively w.r.t. the operator $\hat{j}$. Similarly to $j$, we have its anti-holomorphic counterpart $\bar{j}=\bar\gamma\bar{a}$ which satisfies the same properties in the anti-holomorphic sector.

We can consider a new ``untwisted'' theory\footnote{
We call it ``untwisted'', since the theory we started with is obtained from it by Witten's topological twist of type B, cf. \cite{Witten_mirror, Hori, Losev_Frenkel}.
} with same field content as before and the deformed stress-energy tensor
\begin{equation}\nonumber
\til{T}:=T-\frac12 \dd j
\end{equation}
With respect to the new stress-energy tensor, the fields change their (holomorphic) conformal weights as follows:

\begin{center}
\begin{tabular}{l|c|c}
field & weight w.r.t. $T$ & weight w.r.t. $\til{T}$ \\
\hline
$J$ & $1$ & $3/2$ \\
$G$ & $2$ & $3/2$ \\
$j$ & $1$ (not primary) & $1$ \\
$\gamma$ & $0$ & $1/2$ \\
$a$ & $1$ & $1/2$
\\ 
$\bar\gamma,\bar{a},b,c$ & $0$ & $0$ 
\end{tabular}
\end{center}

Thus, fields $\gamma,dz\,a$ become (even) Weyl spinors $(dz)^{\frac12}\gamma,(dz)^{\frac12}a$ in the untwisted theory. Similarly, $\bar\gamma, d\bar{z}\,a$ become even spinors $(d\bar{z})^{\frac12}\bar\gamma,(d\bar{z})^{\frac12}\bar{a} $.
The fields $b,c$ are unchanged.
Thus, the bundle of basic fields, replacing (\ref{BRST bundle}) in the untwisted theory, is
\begin{equation}\nonumber
\til{\F}=
\underbrace{K^{\otimes \frac12}}_{a}\oplus \underbrace{K^{\otimes \frac12}}_{\gamma}\oplus 
\underbrace{\bar{K}^{\otimes \frac12}}_{\bar{a}}\oplus \underbrace{\bar{K}^{\otimes \frac12}}_{\bar\gamma}\oplus
\underbrace{\underline{\Pi\RR^2}}_{b,c}
\end{equation}
with $K=(T^{1,0})^*\Sigma$ and $\bar{K}=(T^{0,1})^*\Sigma$ the canonical and anti-canonical line bundles on $\Sigma$.

Note that $j$ was not a primary field w.r.t. $T$, due to the $3$-rd order pole in (\ref{Tj OPE}). However, $j$ is a primary field of weight  $(1,0)$ in the untwisted theory. It is a conserved $U(1)$-current and its Fourier modes generate a Heisenberg Lie algebra due to (\ref{jj OPE}).

One obtains the following OPE of the untwisted stress-energy tensor with itself:
\begin{equation}\nonumber
\til{T}(w)\,\til{T}(z)\sim \frac{-3/2}{(w-z)^4}+\frac{2\,\til{T}_z}{(w-z)^2}+\frac{(\dd \til{T})_z}{w-z}+\reg
\end{equation}
Thus, the untwisted theory has central charge $\s{c}=-3$.\footnote{
Forgetting about the BRST structure, we can regard the action (\ref{S coord}) as a superposition of three non-interacting theories: the second-order ghost system (the $bc$ system) with holomorphic/anti-holomorphic central charges $\s{c}_{bc}=\bar{\s{c}}_{bc}=-2$  and two first order chiral systems with Lagrangians $\gamma \ddelb a$ and $\bar\gamma\ddel  \bar{a}$ with central charges 
$\s{c}_{\gamma a}=-1,\bar{\s{c}}_{\gamma a}=0$ and $\s{c}_{\bar\gamma\bar a}=0,\bar{\s{c}}_{\bar\gamma \bar a}=-1$ respectively (in the \emph{untwisted model}, where $a,\gamma,\bar{a},\bar\gamma$ are even spinors). Thus the total central charge of the system is $\s{c}=\bar{\s{c}}=(-2)+(-1)+(0)=-3$. In the topological (twisted) model, central charges of the first-order systems change to $\s{c}_{\gamma a}=2,\bar{\s{c}}_{\gamma a}=0$ and $\s{c}_{\bar\gamma \bar a}=0, \bar{\s{c}}_{\bar\gamma \bar a}=2$, while the central charge of the ghost system remains $-2$. Thus, $\s{c}^\mr{top}=\bar{\s{c}}^\mr{top}=(-2)+(2)+(0)=0$.
} 
In $N$-component theory, the central charge becomes $\s{c}=-3N$.

Consider the Fourier modes of the fields $\til{T},J,G,j$, defined via
\begin{multline}\nonumber
\til{T}_w=\sum_n (w-z)^{-n-2}\til{L}_n,\\
J_w=\sum_r (w-z)^{-r-\frac32}J_r,\;\;
G_w=\sum_s (w-z)^{-s-\frac32} G_s,\;\;
j_w=\sum_p (w-z)^{-p-1} j_p
\end{multline}
with $n,p\in\ZZ$ and $r,s\in \ZZ+\frac12$ for periodic (Neveu-Schwarz) boundary conditions on fermions and $r,s\in\ZZ$ for anti-periodic (Ramond) boundary conditions. These Fourier modes  satisfy the relations of $\mc{N}=2$ superconformal algebra with central charge $\s{c}=-3$:
\begin{multline}\label{N=2 Virasoro Lie brackets}
[\til L_n,\til L_m]=(n-m)\,\til L_{n+m}-\frac14 (n^3-n)\,\delta_{n,-m},
\quad [j_p,j_q]=-p\, \delta_{p,-q},\\
[\til L_n,J_r]=\left(\frac{n}{2}-r\right)J_{n+r},\quad 
[\til L_n,G_s]=\left(\frac{n}{2}-s\right)G_{n+s},\quad
[\til L_n,j_p]=-p\, j_{n+p},\\
[J_r,G_s]=\til L_{r+s}+\frac{r-s}{2}\,j_{r+s}-\frac12 \left(r^2-\frac14\right)\,\delta_{r,-s},\quad
[j_p,J_r]=J_{p+r},\quad [j_p,G_s]=-G_{p+s},\\
[J_r,J_s]=0,\quad [G_r,G_s]=0
\end{multline}
plus the conjugate relation for the Fourier modes of $\til{\bar T},\bar{J},\bar{G},\bar{j}$. Lie brackets, involving one generator from holomorphic sector and one from anti-holomorphic sector, vanish. In the case of $N$-component theory, the three central extension terms in the commutation relations (\ref{N=2 Virasoro Lie brackets}) 
-- those proportional to $\delta_{\bullet,-\bullet}\cdot 1$ -- get multiplied by $N$.

Thus, the fields $\til{T},J,G,j$ together with their anti-holomorphic counterparts define on the untwisted abelian $BF$ theory the structure of an $\mc{N}=(2,2)$ superconformal theory with supersymmetry currents $J$, $G$  and with $j$ the $R$-symmetry current (plus the conjugates). 

\subsubsection{Dictionary between abelian $BF$ theory and the B model}\label{sec: dictionary}
Recall (see e.g. \cite{Witten_mirror,Hori}) that the free $\mc{N}=(2,2)$ supersymmetric sigma model (or Landau-Ginzburg model with zero superpotential) with target $\CC^N$ is defined by the action
\begin{equation}\label{S SUSY sigma model}
S =
2t\int_\Sigma 
d^2 x
\,\Big(\bar\phi_k\,  \ddel \ddelb \phi^k -i\, \bar\psi_{+k}\, \ddelb \psi_+^k -i\, \bar\psi_{-k}\, \ddel \psi_-^k\Big)
\end{equation}
with scalar fields $\phi^k, \bar\phi_k$ 
corresponding to 
the holomorphic and anti-holomorphic coordinates on the target $\CC^N$, respectively, and with $\psi_\pm^k, \bar{\psi}_{\pm k}$ fermions of spin $1/2$; $t$ is a coupling constant (which is irrelevant in the free theory as it can be absorbed into the normalization of fields). Here, bar/no bar on fields corresponds to anti-holomorphic/holomorphic directions on the target and $\pm$ corresponds to holomorphic/anti-holomorphic directions on the source $\Sigma$. In the B-twisted sigma model the action is the same, however fields $\psi_\pm^k$ attain spin $1$ and $\bar\psi_{\pm k}$ attain spin $0$.

Comparing (\ref{S SUSY sigma model}) with (\ref{S coord}),  we see that the $N$-component abelian $BF$ theory (i.e. with coefficient space $\RR^N$) is the B-twisted supersymmetric sigma model with \emph{odd} target $\Pi \CC^N$:
$$
\begin{CD}\mbox{abelian $BF$ theory} 
\\ @A{\mr{twist}}AA  \\ \mbox{untwisted abelian $BF$ theory}
\end{CD}
\xLongrightarrow{\mr{parity\, reversal + complexification}}
\begin{CD} \mbox{B model} \\ @AA\mr{twist}A  \\ \mbox{supersymmetric sigma model}
\end{CD}
$$

We have the following dictionary (we use the notations of \cite{Witten_mirror,Hori} for the B side).
\begin{center}
\begin{tabular}{c|c}
$BF$ theory & B model  \\
\hline 
coefficient space $V=\RR^N$ & target $X=\CC^N$ \\
$c^k$ & $\phi^k$ \\
$b_k$ & $\bar\phi_k$ \\
$a^k$ & $\psi_+^k$ \\
$i\gamma_k$ & $\bar\psi_{+k}$ 
\\
$\bar{a}^k$ & $\psi_-^k$ \\
$i\bar\gamma_k$ & $\bar\psi_{-k}$ \\
$-\frac12 B_k=\frac{i}{2}(\gamma_k-\bar\gamma_k)$
& 
$\theta_k=\frac12 (\bar\psi_{+k}-\bar\psi_{-k})$
\\
$i\lambda_k=i(\gamma_k+\bar\gamma_k)$
& 
$\bar\eta_k=\bar\psi_{+k}+\bar\psi_{-k} \;\; =``d\bar\phi_k"$ \\
$A^k=dz\, a^k+d\bar{z}\, \bar{a}^k$ & $\rho^k=dz\, \psi_+^k+d\bar{z}\, \psi_-^k$ \\
term $\lambda_k\frac{\dd}{\dd b_k}$ in $Q$
& Dolbeault differential on the target $\bar\eta_k\frac{\dd}{\dd \bar\phi_k}$\\
\hline
\multicolumn{2}{c}{$0$-observables} \\
$\bb{O}_z=T_\mr{poly}(\Pi V)\quad = \CC[c^k,B_k]$ & $\oplus_{p,q} H^{0,p}(X,\wedge^q T^{1,0}X)\quad=\CC[\phi^k,\theta_k]$ \\
\hline
\multicolumn{2}{c}{Supercurrents (in untwisted models)} \\
$J$ & $\bar{G}_+$ \\
$\bar{J}$ & $\bar{G}_-$ \\
$G$ & $G_+$ \\
$\bar G$ & $G_-$ \\
\hline 
total $U(1)$-current $j_\tot=dz\,j+d\bar{z}\,\bar j$ & axial R-symmetry current $J_A$
\end{tabular}
\end{center}

\end{document}